\documentclass[conference]{IEEEtran}
\IEEEoverridecommandlockouts                              
\usepackage[cmex10]{amsmath}
\usepackage{amsthm}
\usepackage{amssymb}
\usepackage{mathrsfs}
\usepackage{graphicx}
\usepackage{float}
\usepackage{array}
\usepackage{epstopdf}
\usepackage{multirow}
\usepackage{cite}
\usepackage{color}
\usepackage[T1]{fontenc}

\begin{document}

\title{COVID TV-UNet: Segmenting COVID-19 Chest CT Images Using Connectivity Imposed U-Net}


\author{Narges Saeedizadeh, Shervin Minaee, Rahele Kafieh, Shakib Yazdani and Milan Sonka, \emph{Fellow, IEEE}
\thanks{Narges Saeedizadeh is with Medical Image and Signal Processing Research Center, Isfahan University of Medical Sciences, Iran (e-mail: saeidynarges@gmail.com).}%
\thanks{Shervin Minaee is with Snap Inc., Seattle, WA, USA (e-mail: sminaee@snapchat.com).}%
\thanks{Rahele Kafieh is with Medical Image and Signal Processing Research Center, Isfahan University of Medical Sciences, Iran (e-mail: rkafieh@amt.mui.ac.ir).}%
\thanks{Shakib Yazdani is with ECE Department, Isfahan University of Technology, Iran (e-mail: shyazdani@ec.iut.ac.ir).}%
\thanks{Milan Sonka is with Iowa Institute for Biomedical Imaging, The University of Iowa, Iowa City, USA (e-mail: milan-sonka@uiowa.edu).}
\thanks{M. Sonka research effort is supported, in part, by NIH grant R01-EB004640.}
}

\maketitle

\begin{abstract}
The novel corona-virus disease (COVID-19) pandemic has caused a major outbreak in more than 200 countries around the world, leading to a severe impact on the health and life of many people globally. As of mid-July 2020, more than 12 million people were infected, and more than 570,000 death were reported.
Computed Tomography (CT) images can be used as an alternative  to  the  time-consuming  RT-PCR  test, to detect COVID-19. 
In this work we propose a segmentation framework to detect chest regions in CT images, which are infected by COVID-19.
We use an architecture similar to U-Net model, and train it to detect ground glass regions, on pixel level. 
As the infected regions tend to form a connected component (rather than randomly distributed pixels), we add a suitable regularization term to the loss function, to promote connectivity of the segmentation map for COVID-19 pixels. 2D-anisotropic total-variation is used for this purpose, and therefore the proposed model is called "TV-UNet". 
Through experimental results on a relatively large-scale CT segmentation dataset of around 900 images, we show that adding this new regularization term leads to 2\% gain on overall segmentation performance compared to the U-Net model.
Our experimental analysis, ranging from visual evaluation of the predicted segmentation results to quantitative assessment of segmentation performance (precision, recall, Dice score, and mIoU) demonstrated great ability to identify COVID-19 associated regions of the lungs, achieving a mIoU rate of over 99\%, and a Dice score of around 86\%.
\end{abstract}

\IEEEpeerreviewmaketitle

\section{Introduction}
Since December 2019, a novel corona-virus (SARS-CoV-2) has spread from Wuhan to the whole China, and then to many other countries. At the end of January 2020, the World Health Organization (WHO) declared that COVID-19  a Public Health Emergency of International Concern \cite{Radio1-1}. 
By July 15 2020, more than 12 million confirmed cases, and more than 570,000 deaths cases were reported across the world \cite{Radio1}. While infection rates are decreasing in some countries, numbers of new infections continue quickly growing in many other countries, signaling the continuing and global threat of COVID-19 \cite{covid_work1, covid_work2, covid_work3}.

Up to this point, no effective treatment has yet been proven for COVID-19. Therefore for prompt prevention of COVID-19 spread, accurate and rapid testing is extremely pivotal. The reverse transcription polymerase chain reaction (RT-PCR) has been
considered the gold standard in diagnosing COVID-19. However, the shortage of available tests and testing equipment in many areas of the world limits rapid and accurate screening of suspected subjects. Even under best circumstances, obtaining RT-PCR test results takes more than 6 hours and the routinely achieved sensitivity of RT-PCR is insufficient \cite{Radio1-2}.
On the other hand, the radiological imaging techniques like chest X-rays and computed tomography (CT) followed by automated image analysis \cite{Minaee-MedIA-2020} may successfully complement RT-PCR testing. 
CT screening provides three-dimensional view of the lung and is therefore more sensitive (although less widely available) compared to chest X-ray radiography. 

In a systematic review \cite{Radio1-5} the authors indicated that CT images are sensitive in detection of COVID-19 before observation of some clinical symptoms. Typical signs of COVID-19 in CT images consist of unilateral, multifocal and peripherally based ground glass opacities (GGO), interlobular septal thickening, thickening of the adjacent pleura, presence of pulmonary nodules, round cystic changes, bronchiectasis, pleural effusion, and lymphadenopathy \cite{Radio1-3,Radio1-4}. 
Accurate and rapid detection and localization of these pathological tissue changes is critical for early diagnosis and reliable severity assessment of COVID-19 infection. As the number of affected patients is high in most of the hospitals, manual annotation by well-trained expert radiologists is time consuming and subject to inter- and intra-observer variability. Such annotation is tremendous and labor-intensive for radiologists and slows down the CT analysis. The urgent need for automatic segmentation of typical COVID-19 CT signatures is widely appreciated and deep learning methods can offer a unique solution for  identifying COVID-19 signes of infection in clinical-quality images frequently suffering from variations in CT acquisition parameters and protocols \cite{Radio1-6}. 

In this work, we present a deep learning based framework for automatic segmentation of pathologic COVID-19-associated tissue areas from clinical CT images available from publicly available COVID-19 segmentation datasets.
There has been a huge progress in the performance of image segmentation model using various deep learning based frameworks in recent years \cite{segnet, deeplab, ccnet, segsurvey, biosurvey, unet}.
Our solution is based on adapting and enhancing a popular deep learning medical image segmentation architecture U-Net to for COVID-19 segmentation task.
As COVID-19 tissue regions tend to form connected regions identifiable in individual CT slices, "connectivity promoting regularization" term was added to the specifically designed training  loss function to encourage the model to prefer sufficiently large connected segmentation regions of desirable properties. 
It is worth to mention that there have been a few works proposed for COVID-19 segmentation from CT images very recently. 
In \cite{infnet}, Fan et al. proposed Inf-Net, to identify infected regions from chest CT slices. 
In they proposed model a parallel partial decoder is used to aggregate the high-level features and generate a global map. Then, the implicit reverse attention and explicit edge-attention are utilized to model the boundaries and enhance the representations. But unfortunately this model used a very small dataset of CT labeled images for segmentation, which consist of a total of 100 CT slices, making it hard to generalize the result, and compare. 
In terms of Dice score, achieves much higher Dice score, on a larger test set.
In \cite{encdec}, Elharrouss et al. proposed an encoder-decoder based  model for lung infection segmentation using CT-scan images. 
The proposed model first uses image structure and texture to extract the ROI of infected area, and then uses the ROI along with image structure to predict the infected region.
They also train this model on a small dataset of CT images, and achieve reasonable performance.
In \cite{3dunet}, Ma et al. prepared a new benchmark of 3D CT data with 20 cases that contains 1800+ annotated slices, and provided several pre-trained baseline
models, that serve as out-of the-box 3D segmentation.

\if false 
When evaluating the performance of our model, both in terms of quantitative and qualitative analysis, .
For quantitative analysis, we looked several metrics, such as precision, recall, Dice score, and mean-intersection-over-union (mIoU), precision-recall curve.
We also provide a comparison of this model, with some powerful baseline model (such as fine-tuned U-Net on this dataset), and show better performance both quantitatively, and qualitatively.
\fi 

The main contributions of our work can be summarized as follows:
\begin{itemize}
    \item Development of an image segmentation framework for detecting pathologic COVID-19 regions in pulmonary CT images,
    \item Development of a novel connectivity-promoting regularization loss function,
    \item Quantitative validation showing 
    better performance than some of existing state-of-the-art segmentation approaches
    \item  Publicly sharing the developed software code facilitating research and medical community use.
\end{itemize}

\begin{figure}[h!]
\begin{center}
   \includegraphics[page=2,width=0.8\linewidth]{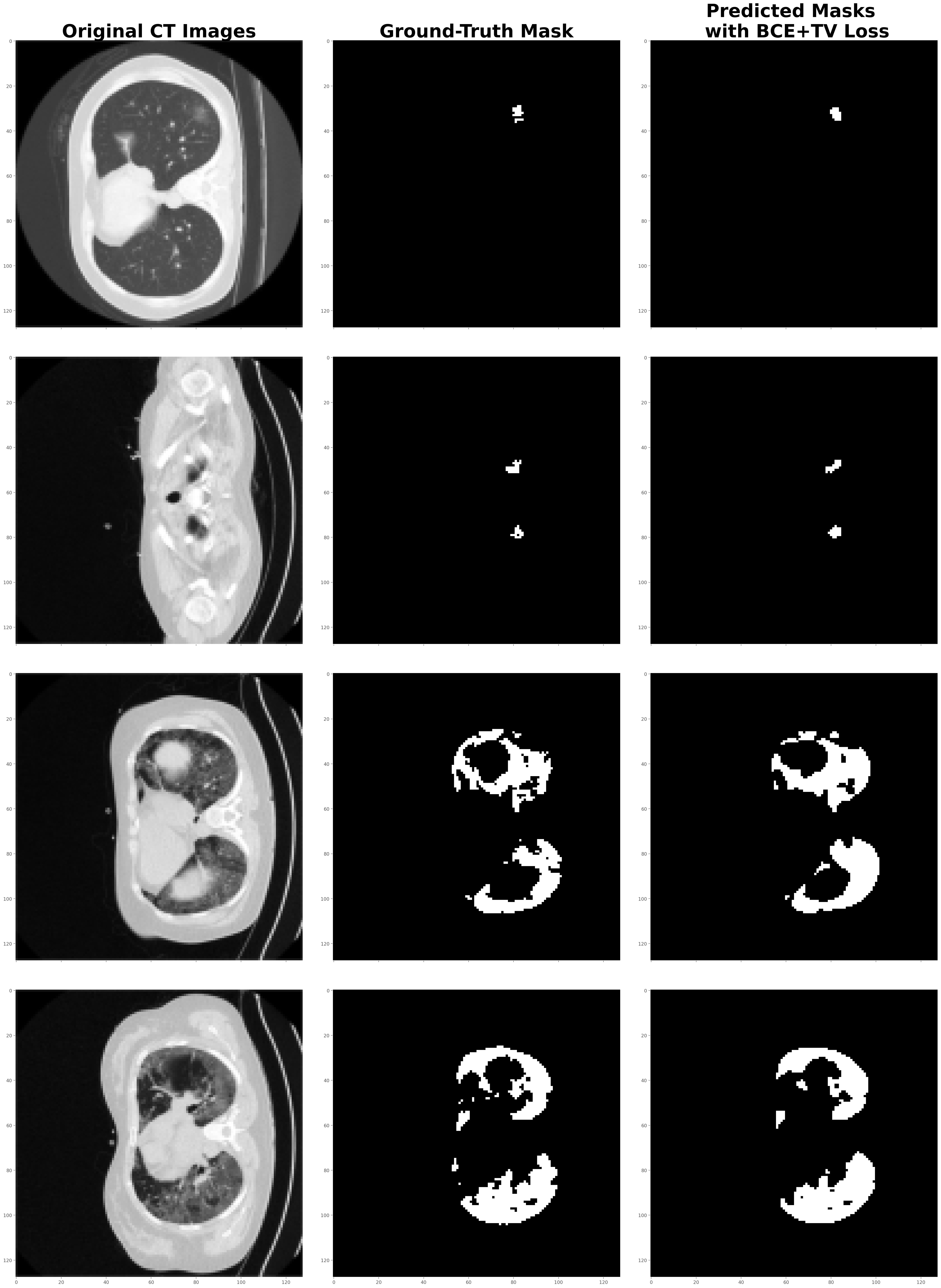}
\end{center}
   \caption{Predicted segmentation masks by the proposed TV-UNet for a few samples images from test.}
\label{result_4mask}
\end{figure}

\section{The Proposed Framework}
\label{sec_method}
Despite a large number of patients suffering from COVID-19, despite a growing number of COVID-19 volumetric CT scans, the availability of labeled CT images that can be used for training of deep learning methods is still limited. Therefore, our strategy relies heavily on the use of transfer learning, initiating the training from a model previously developed for  medical image segmentation (segmentation of neuronal structures in electron  microscopic stacks), and adapt it toward this task.
To better suit the segmentation task at hand, we employed an architecture  similar to the U-Net, one of the most successful deep learning  medical image segmentation approaches, and modified its loss function to prefer COVID-19 specific foreground mask connectivity.

\subsection{U-Net Architecture}
\begin{figure}[ht]
\begin{center}
   \includegraphics[page=2,width=0.99\linewidth]{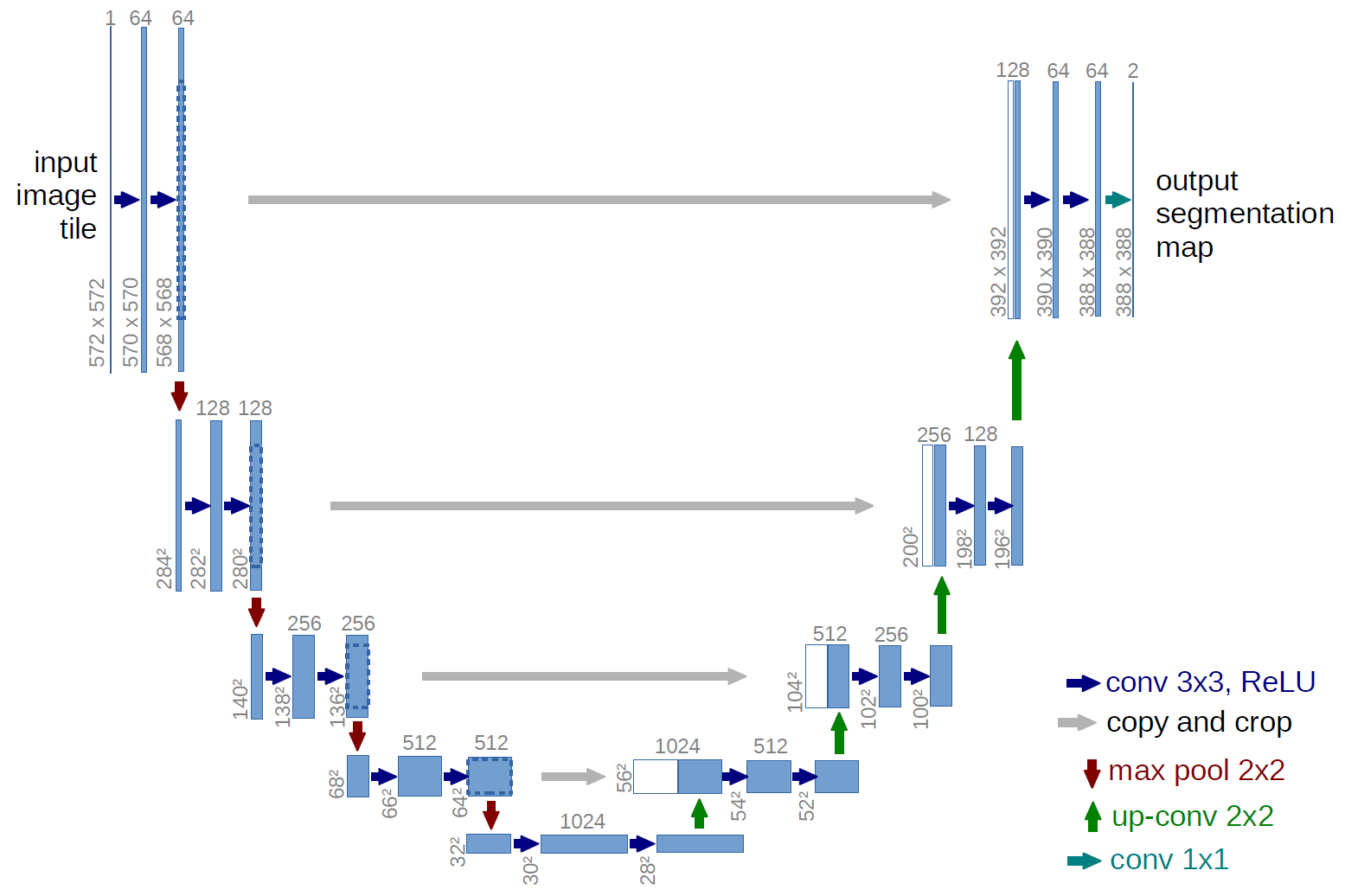}
\end{center}
   \caption{The architecture of U-Net model}
\label{fig:unet}
\end{figure}

U-Net is one of the popular segmentation models which is based on encoder-decoder neural architecture and use of skip connections, and was originally proposed by Ronneberger et al. \cite{unet}. 
The network architecture of U-Net is illustrated in Fig.\ \ref{fig:unet}. 
In the encoder part, model gets an image as input and applies multiple layers of convolution, max-pooling and ReLU activation, and compresses the data into a latent space.
In the decoder part, the network attempts to decode the information from the latent space using transposed convolution operation (deconvolution) and produce the segmentation mask of the image. The rest of the operations are similar to the aforementioned ones in the encoder part.
One difference between U-Net and plain encoder-decoder model is the use of skip-connections to send the information from the corresponding high-resolution layers of the encoder to the decoder, which can help the network to better capture small details that are present in high-resolution.
Fig.\ \ref{fig:unet} illustrates the general architecture of a U-Net model.

Similar to other neural segmentation models, U-Net uses the loss function in Eq \ref{unet_loss} during training. 
It is comprised of the weighted cross entropy loss together with the soft-max function over the final feature maps.
\begin{equation}
\begin{aligned}
&  \mathcal{L}_{UNet}=  \sum\limits_{x \in \Omega}\omega(x) \log(p_{\ell(x)}(x)) \;\; ,
\end{aligned}
\label{unet_loss}
\end{equation}
Where $\ell$ : $\Omega$ $\rightarrow$ \{1, . . . , K\} considering that $\Omega$ $\subset$ $\mathbb{Z}^2$ and also K denotes the total number of classes in the dataset. Moreover, the soft-max is defined as $p_{k}(x)$ = $\exp(a_{k}(x))$ $\slash$ $\sum \limits_{k'=1}^K\exp(a_{k'}(x))$ where $a_{k}(x)$ represents the activation in feature channel K.  Additionally, $\omega $ : $\Omega$ $\rightarrow$  ${\rm I\!R}$ is a weighted map to give some features more importance.


\subsection{Connectivity Regularization}
The segmentation maps usually consist of a number of connected components, and single-pixel regions are rare. 
To encourage our segmentation model to generate segmentation maps with connected components of desirable sizes, we found that incorporating an explicit regularization term in the training loss function can greatly improve connectivity requirements for the predicted segmentation regions. 
It is worth noting that conventional U-Net can also implicitly learn such behavior from  training data to some extent, assuming sufficient data sizes are available, which is not quite the case in our situation.
Several strategies were developed and considered to impose desired connectivity requirements within images such as adding group-sparsity or incorporate total variation terms \cite{TV, TV2, group-sparse}. 
Based on achieved experience, we decided to use total variation (TV)  as its gradient update is computationally attractive during the backward pass stage.

TV penalizes the generated images with large variations among neighboring pixels, leading to more connected and smoother solutions \cite{TV2}. Total variation of a differentiable function $f$ defined on an interval $[a,b] \subset R$ has the following expression if $f'$ is Riemann-integrable:
\begin{equation}
\begin{aligned}
V_a^b= \int_{a}^{b} \| f'(x) \| dx \;\; .
\end{aligned}
\end{equation}

Total variation of 1D discrete signals ($y=[y_1,  ..., y_N]$) is straightforward, and can be defined as:
\begin{equation}
\begin{aligned}
TV(y)=  \sum_{n=1}^{N-1} |y_{n+1}-y_{n}| \;\; ,
\end{aligned}
\end{equation}
where $D$ is a $(N-1)\times N$ matrix as below:
\begin{equation*}
D= 
\begin{bmatrix}
    1       & -1 & 0 & \dots & 0 \\
    0       & 1 & -1 & \dots & 0 \\
    \vdots & \vdots & \vdots & \ddots & \vdots \\
    0       & 0 & \dots &  1 & -1
\end{bmatrix} \;\; .
\end{equation*}

For 2D signals ($Y=[y_{i,j}]$), we can use isotropic or  anisotropic versions of 2D total variation \cite{TV}. 
To simplify our optimization problem, we have used the anisotropic version of TV, which is defined as the sum of horizontal and vertical gradients at each pixel:
\begin{equation}
\begin{aligned}
TV(Y)=  \sum_{i,j} |y_{i+1,j}-y_{i,j}|+|y_{i,j+1}-y_{i,j}| \;\; .
\end{aligned}
\end{equation}

In our case we can add the total variation of the predicted binary mask for COVID-19 pixels to the loss function.
Adding this 2D-TV regularization term to our framework will promote the connectivity of the produced segmentation regions. 
The new loss function for our model would then be defined as:
\begin{equation}
\begin{aligned}
&  \mathcal{L}_{TV-UNet}=  \mathcal{L}_{UNet}+ \lambda \ TV(M(x)) \;\; ,
\end{aligned}
\end{equation}
where $\mathcal{L}_{Unet}$ is the binary cross-entropy loss, which is similar to the sum of cross-entropy on all pixels, and defined above.

\section{COVID CT Segmentation Dataset}
\label{sec_dataset}
We have used the \textbf{COVID-19 CT segmentation dataset} \cite{Radio9}, which contains two versions. The first version
of this dataset contains 100 images from 40 patients, which are all labeled as COVID-19 class. This dataset has
three types of ground truth masks, which are called Ground
Glass, Consolidation and Pleural Effusion. The original CT images and all ground truth masks have a size of 512 x 512.
The second version of the dataset was
expanded to 829 images (from 9 patients) in which 373 of those are labeled as COVID-19 and the rest as normal. 
All ground truth masks as COVID-19 have Ground
Glass mask but majority of them are missing the latter two. The size of images and masks in the second version of this dataset is 630 x 630. 
We combined these two versions, which contains a total of 49 people.

\begin{figure}[h]
\begin{center}
   \includegraphics[page=2,width=0.99\linewidth]{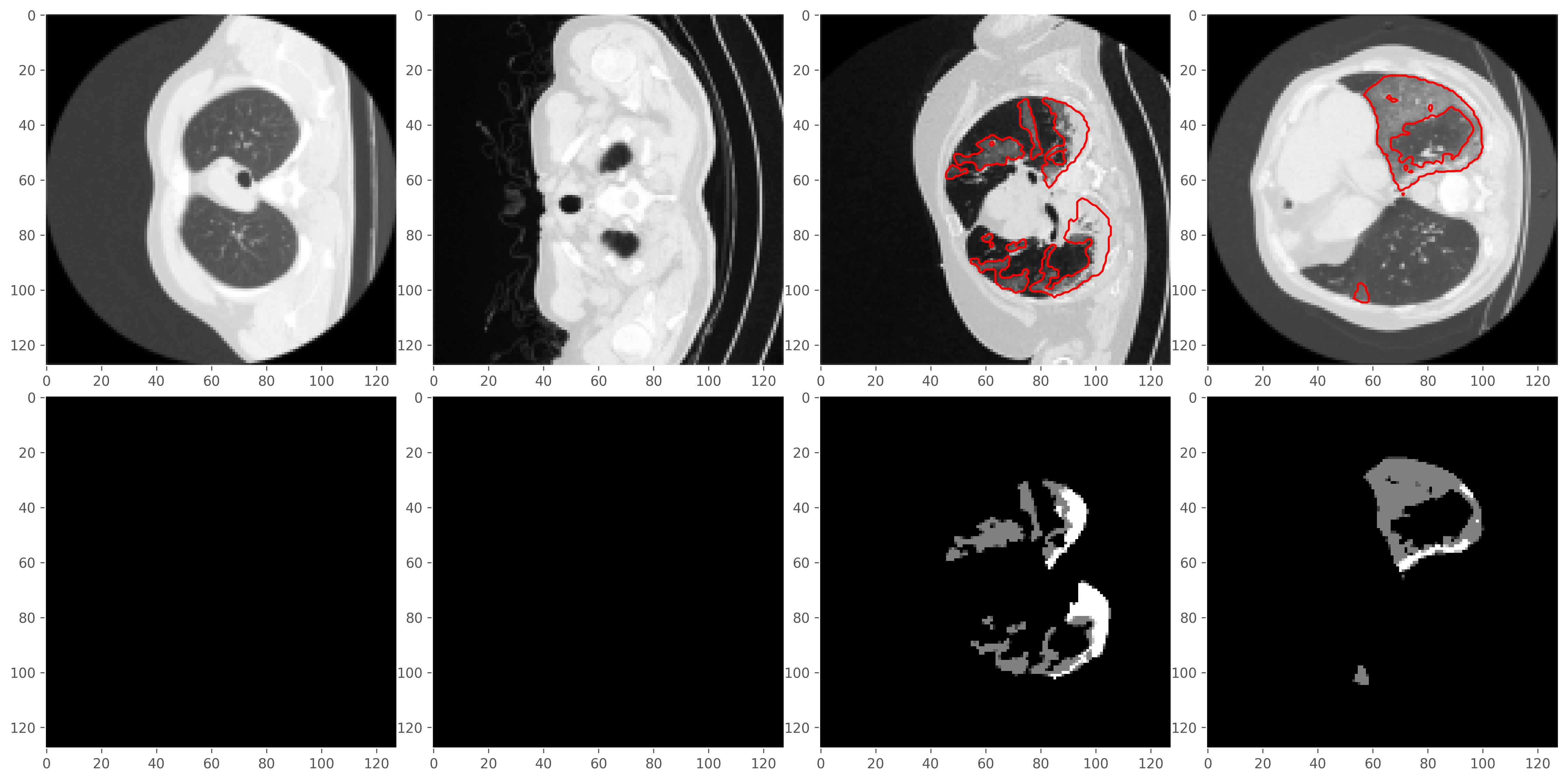}
\end{center}
   \caption{Four sample images from the COVID-19 CT segmentation  dataset. The first two columns show two normal images, and the images in the third and fourth columns deote COVID-19 Images. 
   There are 929 images in this dataset, from which 473 images are labeled as COVID-19, and 456 as healthy.}
\label{fig:NORMAL}
\end{figure}

Four sample images from this dataset are shown in Figure \ref{fig:NORMAL}.
The images in the first and second rows denote the original images, and their corresponding COVID-19 mask, respectively.
The images in the first and second columns denote two sample images of normal people, and the images in the third and fourth columns denote two COVID-19 images.
The white and gray regions in ground-truth masks denote COVID-19 regions, while the black pixels denote healthy regions (note that if the mask is entirely black, it means that the given CT image belongs to a healthy person). 
The red boundary contours are drawn to better show the parts containing COVID-19, and are not a part of the original image.

\begin{figure}[h!]
  \includegraphics[page=2,width=0.99\linewidth]{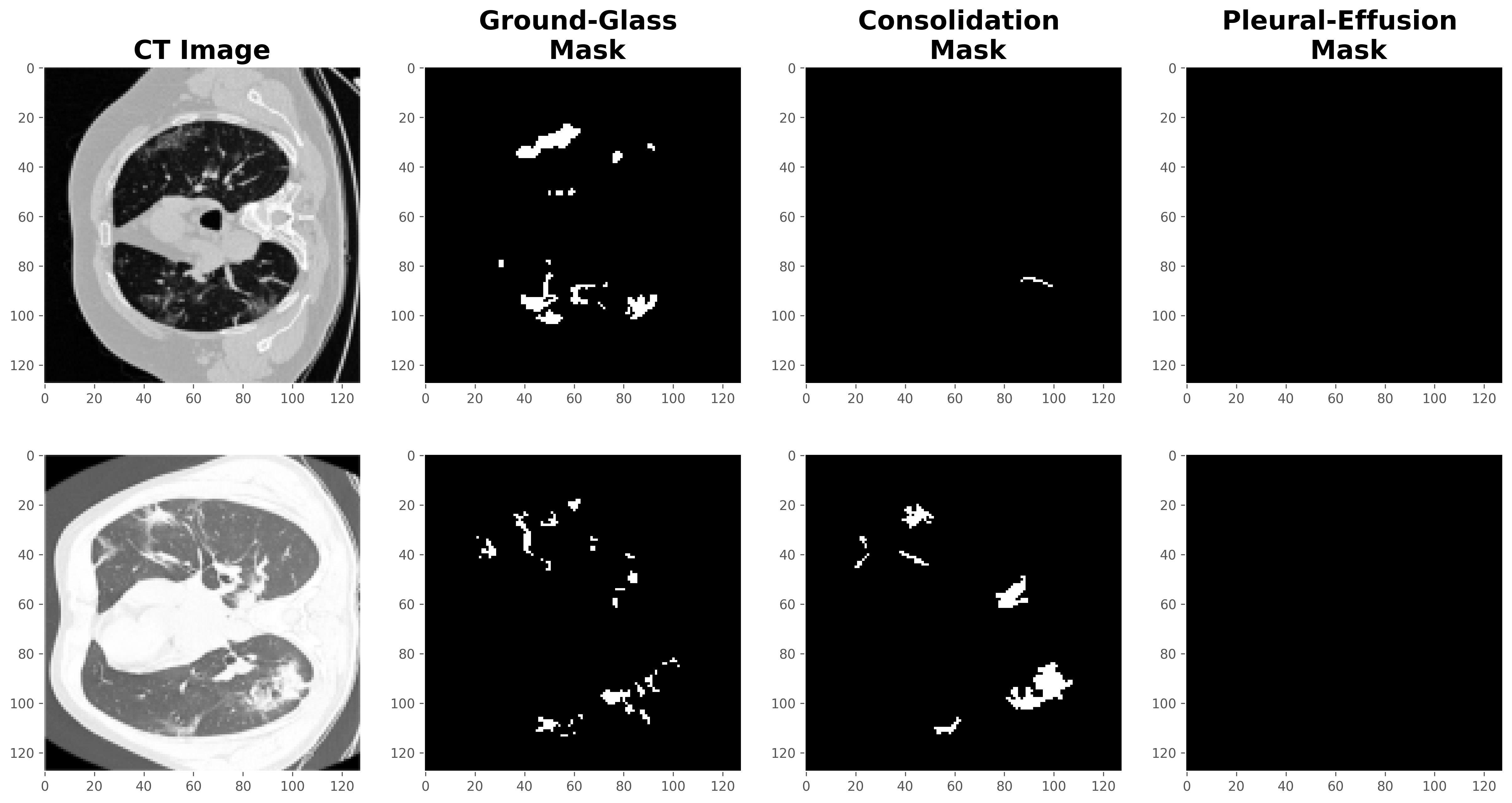}
  \caption{CT Images with three types of Ground-Truth Mask}
  \label{fig:mask_types}
\end{figure}

After precise examination of the three types of ground truth masks, and consulting with a board-certified radiologist, we decided to focus on the Ground Glass mask, and remove the Consolidation and Pleural Effusion masks, as:
on one hand very few images have all three types of masks and most of them are missing the latter two, and on the other hand it is verified by our radiologist that the result using only ground-glass mask is also acceptable and it can be used to infer the presence of COVID-19. 
Two sample images of COVID-19 class with the three types of ground truth masks are shown in Figure \ref{fig:mask_types}.

\subsection{Training and Test Split}
To evaluate the effect of different ground truth masks and the effect of different training/testing set composition, two different splits (of training, validation, and test sets) are selected from this dataset. 
In split-1, we have 729 images (associated with 46 patients) in the train and validation sets, and 200 images (associated with 3 patients) in the test set.  
In split-2, we have 654 images (associated with 35 patients) in the train and validation sets, and 275 images (associated with 14 patients) in the test set. 
The details of these splits are provided in Table \ref{tab:dataset_split}.

\begin{table}[ht]
\centering
  \caption{Train/Validation/Test of the final dataset without Data Augmentation for different split}
  \centering
\begin{tabular}{|l|c|c|}
\hline
Data  &  Number of Images in Split-1 &  Number of Images in Split- 2 \\
\hline
Training  & 654 & 590\\ \hline
Validation & 75&64\\\hline
Test  & 200& 275\\ \hline
Total  & 929 &929\\ \hline
\end{tabular}
\label{tab:dataset_split}
\end{table}

Additionally,  a semi-supervised COVID-19  segmentation dataset  (COVID-SemiSeg) recently reported in  \cite{infnet} and  \cite{Semi-Seg} was used to compare our TV-Unetapproach  with other methods. The COVID-SemiSeg dataset consists of two sets. The first one contains 1600 pseudo labels generated by  Semi-Inf-Net model and 50 labels by expert physicians. The second set included 50 multi-class labels. There are 48 images to used for performance assessment in both sets.

\section{Experimental Results}
\label{sec_result}
In this section we provide a detailed experimental analysis of the proposed segmentation framework, by presenting both qualitative and quantitative results as well as comparing our results with a baseline approach.

\subsection{Evaluation Metrics }
There are several metrics that are used by the research community to measure the performance of segmentation models, including precision, recall, dice coefficient and mean IoU (mIoU).
These metrics are also widely used in medical domain, and are defined as below.

\textbf{Precision} is calculated as the ratio of pixels correctly predicted as COVID divided by total pixels predicted as COVID, and is defined as Eq. \ref{pre}:
\begin{equation} \label{pre}
\begin{aligned}
Precision & = \frac{TP}{TP+FP} \;\; , 
\end{aligned}
\end{equation}
where TP refers to the true positive (the number of correctly predicted COVID-19 cases) and FP refers to the false positive (the number of wrongly predicted COVID-19 cases).

\textbf{Recall} is the ratio of pixels correctly predicted as COVID-19 divided by total number of actual COVID-19 pixels, and is defined as Eq. \ref{Recall}:
\begin{equation} \label{Recall}
\begin{aligned}
Recall & = \frac{TP}{TP+FN} \;\; , 
\end{aligned}
\end{equation}
where TP is false positiv, and FN refers to the false negative and is the number of pixels mistakenly predicted as non-COVID.

Precision and Recall are widely used in medical domain, and to get a big picture of model performance usually a paired version of them is used.  
Precision-Recall (PR) curve is popular way to look at the model performance holistically, which is a plot of the precision (y-axis) versus
the recall (x-axis) rates for different thresholds.

\textbf{Dice Coefficient} (also known as Dice score, or DSC) is another popular metric especially for the multi-class image segmentation. 
The dice score is defined as Eq.\ \ref{DSC}:
\begin{equation} \label{DSC}
\begin{aligned}
Dice& = \frac{2|A\cap B|}{|A|+|B|}\\
\end{aligned} \;\; ,
\end{equation}
where A and B denote the predicted and ground-truth masks. 

\textbf{Intersection over Union} (also known as Jaccard index) is another popular metric used to evaluate the similarity between ground truth and predicted segmentation masks. 
It is defined as the size of the intersection divided by the size of the union of the target mask and predicted segmentation map (Eq. \ref{IOU}).

\begin{equation} \label{IOU}
\begin{aligned}
IoU(A,B) & = \frac{|A\cap B|}{|A \cup B|} \\
\end{aligned} \;\; ,
\end{equation}
where A and B are predicted and  ground-truth masks. If A and B are both empty, IoU(A,B) is defined as 1. 
IoU ranges between 0 and 1.
\textbf{Mean-IoU} is the average IoU values over all classes. 
It is worth mentioning that Dice coefficient and IoU are positively correlated.

\subsection{Model Hyper-parameters}
Hyper-parameters are very important, and it is crucial to properly tune their values during the training of machine learning models to achieve good performance, especially in the case of deep neural networks.
Hyper-parameter tuning can be done in two different ways, automatically and manually. 
In this work, we manually evaluated few different combinations of  hyper-parameters and selected the best combination. 
To simplify the tuning process, we fixed the number of epochs to 100, and the batch-size to 32.
We designed and compared different loss functions (such as binary cross entropy (BCE), dice coefficient loss, and BCE plus total variation regularization), different optimizers (such as ADAM, Adagrad, Adadelta and stochastic gradient descent (SGD)), and different learning rates.
We used adaptive learning rate scheduling and early stopping criteria as below, which achieved reasonable performance on the validation set: 
\begin{itemize}
    \item Learning rate is decayed whenever the validation loss does not improve for 5 continues epochs.
    \item Early stopping is applied whenever the validation loss does not improve for 10 subsequent epochs.
\end{itemize}

Table \ref{tab:Different Losses} shows the impact of the loss function design on the model performance with binary cross entropy and the proposed connectivity regularized loss function achieving the best performance.

 \begin{table}[h!]
\centering
  \caption{Overall performance with different Loss Functions employed, the best cut-off threshold of 0.3 used.
  Best performance shown in bold font.
  }
 \centering
\begin{tabular}{ |p{1.50cm}|p{0.99cm}|p{0.8cm}|p{0.7cm}| p{0.7cm}|p{1.025cm}| }
 \hline
Loss &	Optimizer  &	Learning  
Rate&	mIOU&	DSC&	Average 
\ \ \ \ \ Precision \\
 \hline
BCE& 	&	&0.993&	0.839&	0.92 
\\
BCE+DSC&	&	&0.993&	0.843&	0.91 
\\
BCE+DSC+TV& {ADAM} &{0.001}	& 0.988& 0.645&	0.91 
\\
BCE+TV&	 &	&	0.995&	\textbf{0.864}&	\textbf{0.94}\\
 \hline
 \end{tabular}
 \label{tab:Different Losses}
\end{table}
 
The impact of the optimizer on the model performance is shown in Table \ref{tab:Different Optimizer}.
As we can see, ADAM achieves the highest performance in terms of all metrics.

 \begin{table}[h!]
\centering
  \caption{Overall  performance for different Optimizer selection, using the best cut-off threshold of 0.3.
  Best performance shown in bold font.
  }
\centering
 \begin{tabular}{ |p{1.10cm}|p{1cm}|p{1cm}|p{0.75cm}| p{0.75cm}|p{1.01cm}| }
 \hline
Loss&	Optimizer&	Learning  
Rate&	mIOU&	DSC&	Average \ \ \ \
Precision\\
 \hline
&ADAM&	&	0.995&	\textbf{0.864}&	\textbf{0.94}\\
&	SGD	& &	0.985&	0.573&	0.8 \\
BCE+TV&	Adadelta& 0.001&	0.991&	0.780&	0.9 \\
 &	Adagrad	& &	0.992&	0.784&	0.9 \\
 \hline
 \end{tabular}
 \label{tab:Different Optimizer}
\end{table}
 
Table \ref{tab:Different Learning Rate} provides the analysis of model performance for two different learning rate values when using (the best performing) ADAM optimization.

\begin{table}[h!]
\centering
  \caption{Overall model performance for different Learning Rates, again for the best cut-off threshold  of 0.3.
Best performance shown in bold font.}
 \centering
 \begin{tabular}{ |p{1.10cm}|p{1cm}|p{1cm}|p{0.75cm}| p{0.75cm}|p{1.01cm}| }
 \hline
Loss&	Optimizer&	Learning  
Rate&	mIOU&	DSC&	Average \ \ \  Precision\\
 \hline
&&0.001	&	0.995&\textbf{0.864}&	\textbf{0.94}\\
  {BCE+TV}&{ADAM} &&&&\\
  &&0.0001&	0.993&	0.838&	0.92\\
 \hline
 \end{tabular}
 \label{tab:Different Learning Rate}
\end{table}


Note that the model predicts a probability for each pixel, showing the likelihood of it belonging to the pathologic  COVID-19 region (zero denotes Non-COVID pixels and one denotes COVID-19 pathology).
These probabilities are  thresholded, different thresholds yield certain sensitivit/specificity rates.
Threshold value of 0.3 achieved the best performance on the validation set and was therefore used to report the results of the proposed model.
The impact of modifying the threshold values on the model accuracy is given in Section \ref{thresholds}.

\subsection{Predicted Masks}
\begin{figure}[h!]
\begin{center}
   \includegraphics[page=2,width=0.99\linewidth]{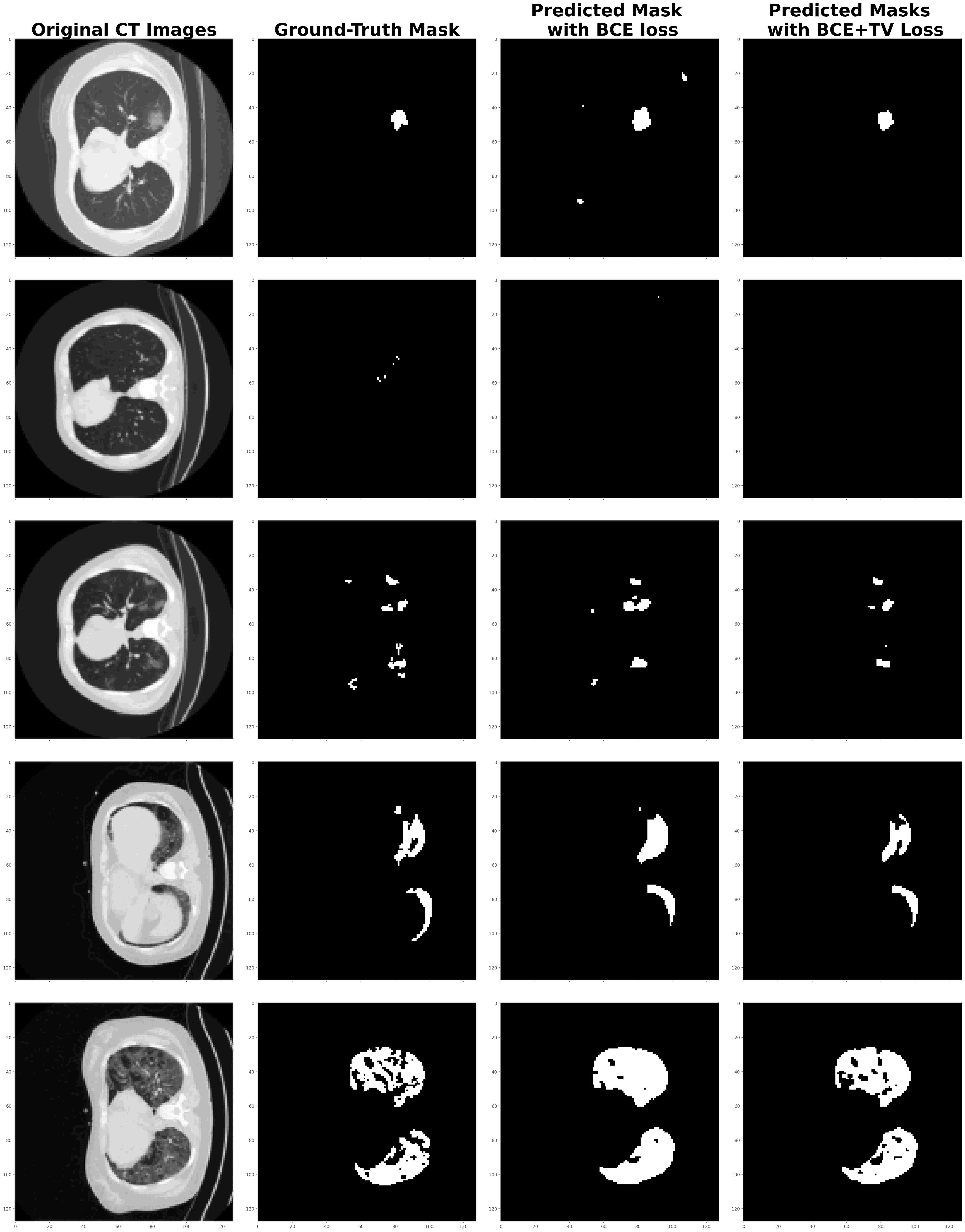}
\end{center}
   \caption{Predicted segmentation masks by conventional U-Net and the proposed TV-UNet for a few samples images from test.
   }
\label{result_5mask}
\end{figure}

Qualitative result showing how close our predicted masks are to the ground-truth masks are given in 
Figure \ref{result_5mask} 
for 5 sample images from the test set. 
As can be seen when the desired region is very tiny, the conventional U-Net model (fine-tuned on our dataset) cannot distinguish the segmentation region and background very well, while the proposed TV-UNet model performs notably better.

\subsection{Cut-off Threshold Impact on Model Performance}
\label{thresholds}
As discussed previously, our model predicts a probability score for each pixel, showing the likelihood of its being in COVID-19 pathology region. Different cut-off thresholds can be used on those probabilities to decide COVID-19 labeling.
By increasing the cut-off threshold, less and less pixels would be labeled as COVID-19 pathology.
Tables \ref{TVUnet_thresh1} and \ref{TVUnet_thresh2} show the model performance (in terms of precision, recall, and mIoU) for eight different values of cut-off thresholds for Split-1  and Split-2 datasets.
The cut-off threshold of 0.3 results in the highest Dice score, and mIoU metric, and therefore was employed to compare our model with other baseline models.
\begin{table}[h!]
\centering
  \caption{Precision, recall, Dice score, and mIoU rates of TV-Unet model for different threshold values for Split 1.
Confidence intervals provided for recall metric. Best performance shown in bold font.
  }
  \centering
\begin{tabular}{|c|c|c|c|c|}
\hline
Threshold  & Recall& Precision & mIoU& DSC\\
\hline
0.1&	0.955$\pm$ 0.028&	0.736&	0.992&	0.831\\ \hline
0.2&	0.913$\pm$ 0.039&	0. 811&	0.994&	0.859\\ \hline
\textbf{0.3}&	0.867$\pm$ 0.047&	0. 859&	\textbf{0.994}&	\textbf{0.863}\\ \hline
0.4&	0.813$\pm$ 0.054&	0. 900&	0.994&	0.854\\ \hline
0.5&	0.746$\pm$ 0.060&	0. 933&	0.993&	0.829\\ \hline
0.6&	0.662$\pm$ 0.065&	0. 959&	0.992&	0.783\\ \hline
0.7&	0.547$\pm$ 0.089&	0. 978&	0.990&	0.702\\ \hline
0.8&	0.362$\pm$ 0.066&	0. 990&	0.986&	0.531\\ \hline
\end{tabular}
\label{TVUnet_thresh1}
\end{table}

\begin{table}[h!]
\centering
  \caption{Precision, recall, Dice score, and mIoU rates of TV-Unet model for different threshold values for Split 2.
 Best performance shown in bold font. }
  \centering
\begin{tabular}{|c|c|c|c|c|}
\hline
Threshold  & Recall& Precision & mIoU& DSC\\
\hline
0.1&	0.892	&0.626      &0.987&0.7363\\ \hline
0.2&	0.833	&0.700		&0.989&0.7609\\ \hline
\textbf{0.3}&	0.781	&0.750		&\textbf{0.990}&\textbf{0.7643}\\ \hline
0.4&	0.730	&0.789		&0.990 &0.7582\\ \hline
0.5&	0.674	&0.825		&0.990 &0.7413\\ \hline
0.6&	0.610	&0.859      &0.990 &0.7139	\\ \hline
0.7&	0.535	&0.890		&0.989 &0.6692\\ \hline
0.8&	0.422	&0.926		&0.987 &0.5801\\ \hline
\end{tabular}
\label{TVUnet_thresh2}
\end{table}

\begin{figure}[h!]
\begin{center}
   \includegraphics[page=2,width=0.95\linewidth]{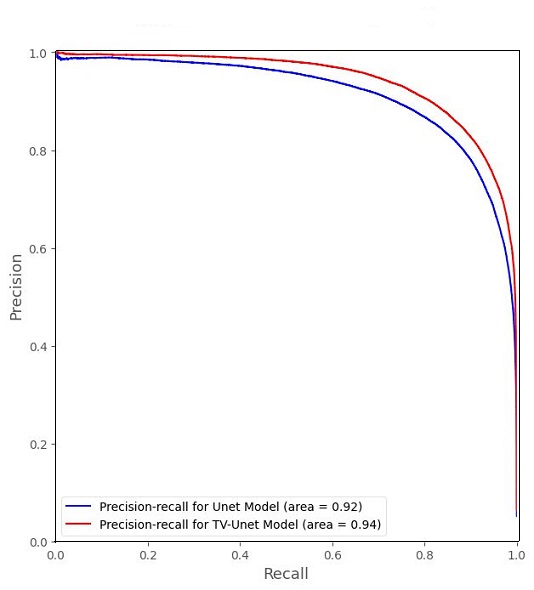}
\end{center}
   \caption{Precision-Recall curve for Split-1.
   }
\label{fig:PR_curve1}
\end{figure}

\begin{figure}[h!]
\begin{center}
   \includegraphics[page=2,width=0.95\linewidth]{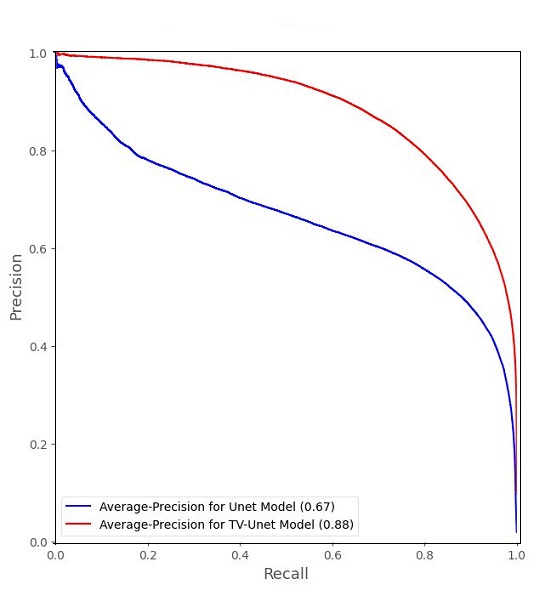}
\end{center}
   \caption{Precision-Recall curve for Split-2. 
   }
\label{fig:PR_curve2}
\end{figure}

To see the holistic view of the proposed model performance on all possible threshold values, Figures \ref{fig:PR_curve1} and \ref{fig:PR_curve2} provide the precision-recall curves on the test sets in Split 1  and Split 2, respectively.
Figure \ref{fig:PR_curve1} shows average precision of 0.92 for the conventional U-Net and 0.94 for our TV-UNet for Split 1 dataset (an improvement of around 0.02 in terms of Average-precision). Figure \ref{fig:PR_curve2} shows average precision of 0.67 for the conventional U-Net and 0.88 for our TV-UNet for Split 2,  a relative improvement of  \textbf{31\%}.

\subsection{Model Performance Comparison with Finetuned U-Net}
For a fair comparison between the proposed TV-Unet model and the conventional fine-tuned U-Net model, corresponding cut-off thresholds were identified for similar recall rates for each model and compared in terms of other performance metrics. 
Tables \ref{tab:Unet and TVUnet compare1} and \ref{tab:Unet and TVUnet compare2} provide the comparison between these two models for four different recall rates. 
Consistency of our TV-UNet model outperforms that of the fine-tuned UNet model when  considering all metrics, 
showing the added  value of the connectivity-promoting regularization. 
We have an average improvement of around 2\% in terms of Dice score in Split 1 and 10.9\% in Split 2 studies.

\begin{table}[h!]
\centering
  \caption{Comparison of conventional U-Net and the proposed TV-UNet model performance in terms of precision, mIOU and DSC for Split 1. 
  Best performance for each method shown in bold font.
  }
  \centering
\begin{tabular}{|c|c|c|c|c|}
\hline
Model  & Recall&Precision&mIOU&  DSC \\\hline
\multirow{4}{4em}{Unet} & 0.975	&0.575&	0.985 &	0.727\\ 
& 0.945	&0.688&	0.990 &	0.798 \\ 
& 0.91&	0.765&	0.992 &	0.832 \\ 
& 0.85&	0.834&	0.993 &	\textbf{0.841} \\ \hline
\multirow{4}{4em}{TV-Unet} &0.975&	0.675&	0.990 &	0.798\\
& 0.945&	0.760&	0.993 &	0.842 \\ 
& 0.91&	0.812&	0. 994 &	0.860 \\ 
& 0.85&	0.871&	0.995 &	\textbf{0.864} \\ \hline
\end{tabular}
\label{tab:Unet and TVUnet compare1}
\end{table}

\begin{table}[h!]
\centering
  \caption{Comparison of the U-Net and TV-UNet model performance in terms of precision, mIOU and DSC for Split 2.
  Best performance for each method shown in bold font.
  }
  \centering
\begin{tabular}{|c|c|c|c|c|}
\hline
Model  & Recall& Precision & mIOU &  DSC \\\hline
\multirow{4}{4em}{Unet} & 0.810&0.594&0.983 &\textbf{0.655}\\ 
&0.643&0.621&0.985 &0.633 \\ 
&0.535&0.655&0.985 &0.595 \\ 
&0.422&0.693&0.984 &0.527 \\ \hline
\multirow{4}{4em}{TV-Unet} &0.810&0.727&0.990 &\textbf{0.764}\\
&0.643&0.842&0.990 &0.729 \\ 
&0.535&0.890&0.989 &0.670 \\ 
&0.422&0.926&0.987 &0.580 \\ \hline
\end{tabular}
\label{tab:Unet and TVUnet compare2}
\end{table}

\subsection{Comparison With Recently-Reported COVID-19 Segmentation Performance}
Quantitative analysis of COVID-19 segmentation performance on CT images is beginning to appear in publications of others.
One such recent model is Inf-Net \cite{infnet}, in which reverse attention mechanism is used in an encoder-decoder based model for COVID-19 segmentation.
This work is trained on COVID-SemiSeg dataset, that was explained in section \ref{sec_dataset}, and tested on a subset of the first version of COVID-CT-segmentation dataset. 
Therefore, for the comparisons in this section our model is trained on COVID-SemiSeg dataset, to have a fair model evaluation setting.

Here we compare the proposed TV-UNet model, with the Inf-Net, and a few promising image segmentation models trained on COVID-SemiSeg dataset, including UNet++ \cite{unet++}, Semi-Inf-Net \cite{infnet}, DeepLab-v3 \cite{deeplabv3}, FCN8s \cite{fcn8s}, and Semi-Inf-Net+FCN8s \cite{infnet}.

Tables \ref{tab:infection-SemiSeg},
\ref{tab:GroungGlass_SemiSeg} and \ref{tab:Consolidation_SemiSeg} provide the performance comparisons in terms of recall and Dice coefficient, in different settings.
As it can be seen from these tables, the proposed TV-UNet model achieves very promising results, outperforming other models in all three experiments with different settings.

\begin{table}[ht]
\centering
  \caption{Comparison of the TV-UNet model performance with  other recent methods in terms of Sensitivity, Specificity and DSC for pathologic regions on COVID-SemiSeg dataset.  Best performance is shown in bold font.}
  \centering
\begin{tabular} { |l|c|c|c|}
\hline
Model  &  Sensitivity & Specificity & Dice Score\\
\hline
U-Net++   & 0.672&0.902&0.518 \\ \hline
Inf-Net  & 0.692&0.943&0.682 \\ \hline
Semi-Inf-Net  & 0.725&0.960&0.739\\ \hline
TV-UNet& \textbf{0.808} & \textbf{0.960} & \textbf{0.801}\\ \hline
\end{tabular}
\label{tab:infection-SemiSeg}
\end{table}

\begin{table}[ht]
\centering
  \caption{Comparison of the TV-UNet model performance with other methods in terms of Sensitivity, Specificity and DSC for the Ground-Glass mask on COVID-SemiSeg dataset. Best performance shown in bold font.}
  \centering
\begin{tabular}{|l|c|c|c|}
\hline
Model  &  Sensitivity & Specificity & Dice Score\\
\hline
DeepLab-v3+ (stride=8)  & 0.478&0.863&0.375 \\ \hline
DeepLab-v3+ (stride=16)   & 0.713&0.823&0.443 \\ \hline
FCN8s &0.537&0.905&0.471\\\hline
Semi-Inf-Net+FCN8s  & 0.720&0.941&0.646\\ \hline
TV-Unet&\textbf{ 0.762}&\textbf{0.979}&\textbf{0.655}\\ \hline
\end{tabular}
\label{tab:GroungGlass_SemiSeg}
\end{table}

\begin{table}[ht]
\centering
  \caption{Comparison of the TV-UNet model performance with several other methods in terms of Sensitivity, Specificity and DSC for the Consolidation mask on COVID-SemiSeg dataset. Best performance shown in bold font.}
  \centering
\begin{tabular}{|l|c|c|c|}
\hline
Model  &  Sensitivity & Specificity & Dice Score\\
\hline
DeepLab-v3+ (stride=8)  & 0.120&0.584&0.117 \\ \hline
DeepLab-v3+ (stride=16)   & 0.245&0.560&0.188 \\ \hline
FCN8s & 0.212 &0.567& 0.221\\\hline
Semi-Inf-Net+FCN8s  & 0.186&0.639&0.238\\ \hline
TV-UNet& \textbf{0.558}&\textbf{0.988}&\textbf{0.537}\\ \hline
\end{tabular}
\label{tab:Consolidation_SemiSeg}
\end{table}

\subsection{Training Convergence Analysis}
\begin{figure}[h!]
\begin{center}
   \includegraphics[page=2,width=0.9\linewidth]{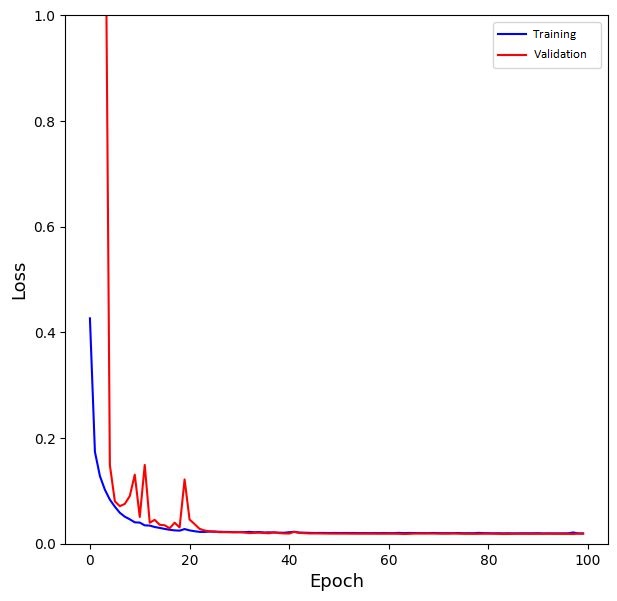}
\end{center}
   \caption{The training and validation loss of the model during training.
   }
\label{training_Loss}
\end{figure}

\begin{figure}[h!]
\begin{center}
   \includegraphics[page=2,width=0.9\linewidth]{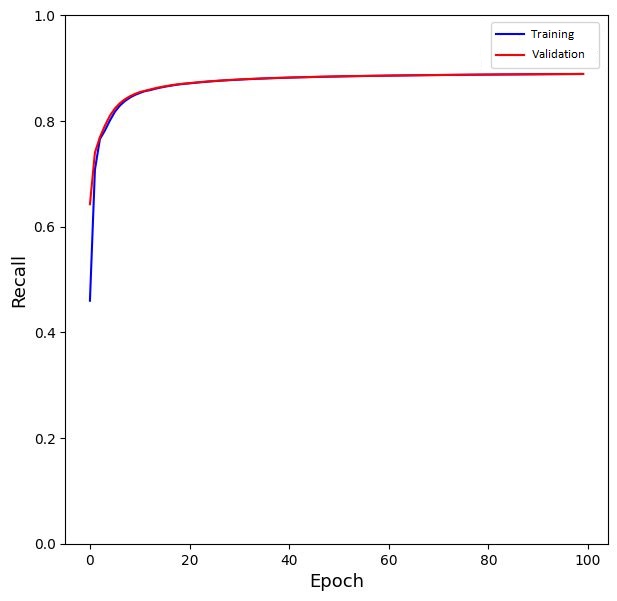}
\end{center}
   \caption{The training and validation recall of the model during training.
    }
\label{training_Recall}
\end{figure}

\begin{figure}[h!]
\begin{center}
   \includegraphics[page=2,width=0.9\linewidth]{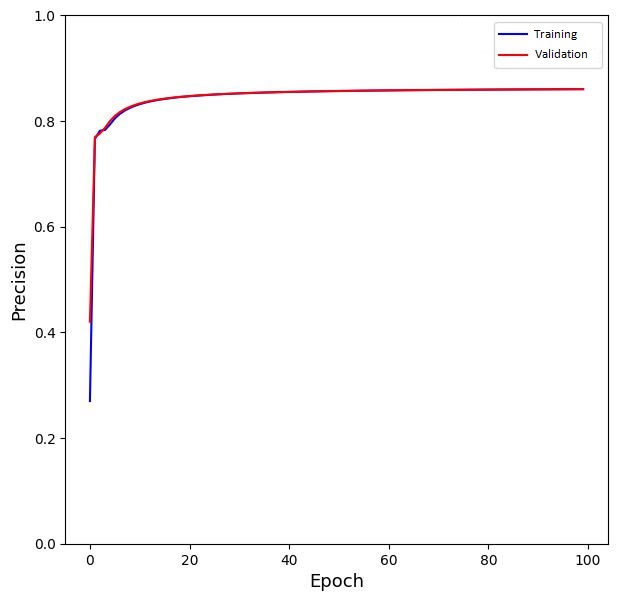}
\end{center}
   \caption{The training and validation precision of the model during training.
       }
\label{training_Precision}
\end{figure}

To see the model convergence during training, we provide the loss function, recall, and  precision rates of the model on different epochs, in Figures \ref{training_Loss}, \ref{training_Recall} and \ref{training_Precision}.
It is worth to mention that for precision and recall, the default threshold value of 0.5 is used in these figures.

\section{Conclusion}
\label{sec_conc}
A novel deep learning framework for COVID-19 segmentation from CT images was reported.
We used the popular U-Net architecture as the main framework, and improved its performance by an added connectivity promoting regularization term, to encourage the model to generate larger contiguous connected segmentation maps.
We showed that the trained model is able to achieve a reasonably high accuracy rate, for detecting of pathologic COVID-19 regions.
We report the model performance under various hyper-parameter settings, which can be helpful for future research by the community to know the impact of different parameters on the final results.
We will further extend this work to semi-supervised setting, in which a combination of labeled and unlabeled data will be used for training the model. 
Such an approach will undoubtedly be extremely useful as collecting accurate segmentation labels for COVID-19 remains very challenging.

\section*{Acknowledgment}
The authors would like to thank our radiologist, Doctor Ghazaleh Soufi, for her advice on the important signals in chest CT images for detecting COVID-19. 
We would also like to thank the providers of the publicly available pulmonary CT datasets. M. Sonka’s research effort supported, in part, by NIH grant R01-EB004640.


\begin{thebibliography}{1}
\small
\bibitem{Radio1}
https://www.worldometers.info/coronavirus/

\bibitem{Radio1-1}
W. H. O. (2020), “Coronavirus disease (COVID-19) pandemic.” https:
//www.who.int/emergencies/diseases/novel-coronavirus-2019.

\bibitem{Radio1-2}
Y. Fang, H. Zhang, J. Xie, M. Lin, L. Ying, P. Pang, and W. Ji,
“Sensitivity of chest CT for COVID-19: Comparison to RT-PCR,”
Radiology, vol. 0, no. 0, p. 200432, 0, pMID: 32073353. [Online].
Available: https://doi.org/10.1148/radiol.2020200432
\bibitem{covid_work1}
Paules, Catharine I., Hilary D. Marston, and Anthony S. Fauci. "Coronavirus infections—more than just the common cold." Jama 323.8 (2020): 707-708.
\bibitem{covid_work2}
Remuzzi, Andrea, and Giuseppe Remuzzi. "COVID-19 and Italy: what next?." The Lancet (2020).
\bibitem{covid_work3}
Kafieh, Rahele, Roya Arian, Narges Saeedizadeh, Shervin Minaee, Sunil Kumar Yadav, Atefeh Vaezi, Nima Rezaei, and Shaghayegh Haghjooy Javanmard. "COVID-19 in Iran: A Deeper Look Into The Future." medRxiv (2020).
\bibitem{Radio1-3}
X. Ding, J. Xu, J. Zhou, and Q. Long, “Chest CT findings of COVID-19 pneumonia by duration of symptoms,” European Journal of Radiology, vol. 127, Jun. 2020, doi: 10.1016/j.ejrad.2020.109009.
\bibitem{Radio1-4}
H. Meng et al., “CT imaging and clinical course of asymptomatic cases with COVID-19 pneumonia at admission in Wuhan, China,” J Infect, Apr. 2020, doi: 10.1016/j.jinf.2020.04.004.
\bibitem{Radio1-5}
S. Salehi, A. Abedi, S. Balakrishnan, and A. Gholamrezanezhad, “Coronavirus
disease 2019 (COVID-19): a systematic review of imaging
findings in 919 patients,” American Journal of Roentgenology, pp. 1–7,2020.
\bibitem{Radio1-6}
F. Shan, Y. Gao, J. Wang, W. Shi, N. Shi, M. Han, Z. Xue, D. Shen, and Y. Shi, “Lung infection quantification of COVID-19 in ct images with deep learning,” arXiv preprint arXiv:2003.04655, 2020.
\bibitem{segnet}
Badrinarayanan, Vijay, Alex Kendall, and Roberto Cipolla. "Segnet: A deep convolutional encoder-decoder architecture for image segmentation." IEEE transactions on pattern analysis and machine intelligence 39.12 (2017): 2481-2495.
\bibitem{deeplab}
Chen, Liang-Chieh, Yukun Zhu, George Papandreou, Florian Schroff, and Hartwig Adam. "Encoder-decoder with atrous separable convolution for semantic image segmentation." In Proceedings of the European conference on computer vision (ECCV), pp. 801-818. 2018.
\bibitem{ccnet}
Huang, Zilong, Xinggang Wang, Lichao Huang, Chang Huang, Yunchao Wei, and Wenyu Liu. "Ccnet: Criss-cross attention for semantic segmentation." In Proceedings of the IEEE International Conference on Computer Vision, pp. 603-612. 2019.
\bibitem{segsurvey}
Minaee, Shervin, Yuri Boykov, Fatih Porikli, Antonio Plaza, Nasser Kehtarnavaz, and Demetri Terzopoulos. "Image segmentation using deep learning: A survey." arXiv preprint arXiv:2001.05566 (2020).
\bibitem{biosurvey}
Minaee, Shervin, Amirali Abdolrashidi, Hang Su, Mohammed Bennamoun, and David Zhang. "Biometric recognition using deep learning: A survey." arXiv preprint arXiv:1912.00271 (2019).
\bibitem{unet}
Ronneberger, Olaf, Philipp Fischer, and Thomas Brox. "U-net: Convolutional networks for biomedical image segmentation." International Conference on Medical image computing and computer-assisted intervention. Springer, Cham, 2015.

\bibitem{infnet}
Fan, Deng-Ping, Tao Zhou, Ge-Peng Ji, Yi Zhou, Geng Chen, Huazhu Fu, Jianbing Shen, and Ling Shao. "Inf-Net: Automatic COVID-19 Lung Infection Segmentation from CT Images." IEEE Transactions on Medical Imaging, 2020.
\bibitem{encdec}
Elharrouss, Omar, Nandhini Subramanian, and Somaya Al-Maadeed. "An encoder-decoder-based method for COVID-19 lung infection segmentation." arXiv preprint arXiv:2007.00861, 2020.
\bibitem{3dunet}
Ma, Jun, Yixin Wang, Xingle An, Cheng Ge, Ziqi Yu, Jianan Chen, Qiongjie Zhu et al. "Towards Efficient COVID-19 CT Annotation: A Benchmark for Lung and Infection Segmentation." arXiv preprint arXiv:2004.12537, 2020.
\bibitem{unet++}
Zhou, Zongwei, Md Mahfuzur Rahman Siddiquee, Nima Tajbakhsh, and Jianming Liang. "Unet++: A nested u-net architecture for medical image segmentation." In Deep Learning in Medical Image Analysis and Multimodal Learning for Clinical Decision Support, pp. 3-11. Springer, Cham, 2018.
\bibitem{deeplabv3}
Chen, Liang-Chieh, Yukun Zhu, George Papandreou, Florian Schroff, and Hartwig Adam. "Encoder-decoder with atrous separable convolution for semantic image segmentation." In Proceedings of the European conference on computer vision (ECCV), pp. 801-818. 2018.
\bibitem{fcn8s}
Long, Jonathan, Evan Shelhamer, and Trevor Darrell. "Fully convolutional networks for semantic segmentation." In Proceedings of the IEEE conference on computer vision and pattern recognition, pp. 3431-3440. 2015.


\bibitem{TV}
A Chambolle, "An algorithm for total variation minimization and applications." Journal of Mathematical imaging and vision 20.1-2: 89-97, 2004.
\bibitem{TV2}
Minaee, Shervin, and Yao Wang. "An ADMM approach to masked signal decomposition using subspace representation." IEEE Transactions on Image Processing 28.7 (2019): 3192-3204.
\bibitem{group-sparse}
Zhang, Jian, Debin Zhao, and Wen Gao. "Group-based sparse representation for image restoration." IEEE Transactions on Image Processing 23.8: 3336-3351, 2014.
\bibitem{Minaee-MedIA-2020}
Minaee, Shervin, Rahele Kafieh, Milan Sonka, Shakib Yazdani, and Ghazaleh Jamalipour Soufi. "Deep-covid: Predicting covid-19 from chest x-ray images using deep transfer learning." arXiv preprint arXiv:2004.09363 (2020).

\bibitem{pytorch}
https://pytorch.org/
\bibitem{Radio9}
http://medicalsegmentation.com/covid19/
\bibitem{Semi-Seg}
https://github.com/DengPingFan/Inf-Net

\end{thebibliography}
\end{document}